\documentstyle[sprocl,psfig]{article}

\input{psfig}
\def\Journal#1#2#3#4{{#1} {\bf #2}, #3 (#4)}

\def\PRP{\em Phys. Rep.}

\def\NPB{{\em Nucl. Phys.} B}
\def\PLB{{\em Phys. Lett.}  B}

\def\PRD{{\em Phys. Rev.} D}

\def\be{\begin{equation}}
\def\ee{\end{equation}}
\def\bea{\begin{eqnarray}}
\def\eea{\end{eqnarray}}

\begin{document}

\title{Electric-Magnetic Duality and the Heavy Quark Potential
      }

\author{ M. BAKER }

\address{Department of Physics,
	 University of Washington,\\
	 Seattle, Washington  98195\\
	 }%


\maketitle\abstracts{We use the assumption of electric-magnetic duality to
express the heavy quark
potential in QCD in terms of a Wilson Loop $W_{\mbox{\scriptsize eff}}(\Gamma)$ determined by the
dynamics of a dual theory which is weakly coupled at long distances.  The
classical approximation gives the leading contribution to $W_{\mbox{\scriptsize eff}}(\Gamma)$
and yields a velocity dependent heavy quark potential which for large $R$
becomes linear in $R$, and which for small $R$ approaches lowest order
perturbative QCD.  The corresponding long distance interaction between color
magnetic monopoles is governed by a Yukawa potential.  As a consequence the
magnetic interaction between the color magnetic moments of the quarks is
exponentially damped.  The semi-classical corrections to
$W_{\mbox{\scriptsize eff}}(\Gamma)$ due to fluctuations of the classical flux tube should lead
to an effective string theory free from the conformal anomaly.
}

\section{Introduction}

In this talk we will describe how to calculate the Wilson loop $W(\Gamma)$
determining the spin dependent, velocity dependent heavy quark potential $V_{q
\bar q}$ using the assumption of electric-magnetic duality; namely, that the
long distance physics of Yang Mills theory depending upon strongly coupled
gauge potentials $A_\mu$ is the same as the long distance physics of a dual 
theory describing the interactions of weakly coupled dual potentials $C_\mu$
and monopole fields $B_i$.  To calculate $V_{q \bar q}$ at long distances we
replace $W(\Gamma)$ by $W_{\mbox{\scriptsize eff}}(\Gamma)$ a functional integral over the
variables of the dual theory \cite{calc}.  Because the long distance fluctuations of the
dual variables are small we can use a semi-classical expansion to evaluate
$W_{\mbox{\scriptsize eff}}$.  The classical approximation gives the dual superconductor picture
of confinement \cite{mand} and the semi-classical corrections lead to an effective string
theory \cite{string}.  We first review electric-magnetic duality in electrodynamics.

\section {Electric-Magnetic Duality in Electrodynamics}

Consider a pair of particles with charges $e(-e)$ moving along trajectories
$\vec z_1(t) (\vec z_2 (t))$
in a relativistic medium having
dielectric constant $\epsilon$.  The trajectories $\vec z_1(t)(\vec z_2(t))$ define world lines $\Gamma_1(\Gamma_2)$ running from
$t_i$ to $t_f(t_f$ to $t_i)$.  The world lines $\Gamma_1(\Gamma_2)$,
along with two straight lines at fixed time connecting $\vec y_1$ to
$\vec y_2$ and $\vec x_1$ to $\vec x_2$, then make up a
closed contour $\Gamma$ (See Fig.1).  The
current density $j^\mu(x)$ then has
the form

\begin{equation}
j^\mu (x) = e \oint_\Gamma dz^\mu \delta (x - z).
\label{eq:2.1}
\end{equation}

\begin{figure}
\leavevmode
\centering
\psfig{file=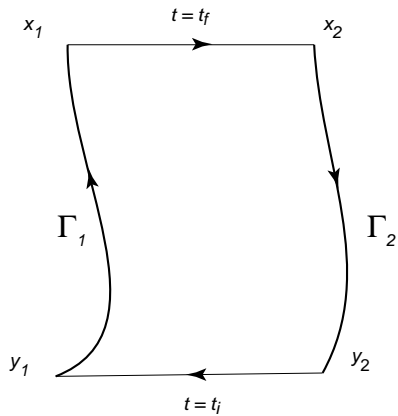,height=2.2in}
\caption{The loop $\Gamma$.}
\end{figure}

In the usual $A_\mu$ (electric) description this system is described by a
Lagrangian
\begin{equation}
{\cal L}_A(j) = - {\epsilon\over 4} (\partial_\alpha A_\beta - \partial_\beta
A_\alpha)^2 - j^\alpha A_\alpha.
\label{eq:2.2}
\end{equation}

\noindent 
Then
\begin{equation}
\int dx {\cal L}_A (j) = - \int dx {\epsilon (\partial_\mu A_\nu -
\partial_\nu A_\mu)^2\over 4} - e \oint_\Gamma dz^\mu A_\mu (z).
\label{eq:2.3}
\end{equation}

\noindent 
The functional integral defining $W(\Gamma)$ in electrodynamics is

\begin{equation}
W(\Gamma)={\int {\cal D} A_\mu e^{i\int dx [{\cal L}_A (j) + {\cal
L}_{GF}]}\over \int {\cal D} A_\mu e^{i\int dx [{\cal L}_A (j = 0) + {\cal
L}_{GF}]}},
\label{eq:2.4}
\end{equation}

\noindent 
where ${\cal L}_{GF}$ is a gauge fixing term.

The spin independent electron positron potential $V_{e^+ e^-} 
(\vec R, {\dot{\vec z}}_1, {\dot{\vec z}}_2 )$ 
is obtained from the expansion of $i$ log
$W(\Gamma)$ to second order in the velocities ${\dot{\vec z}}_1$ and
${\dot{\vec z}}_2$ by the equation:

\begin{equation}
i \log W(\Gamma) = \int_{t_{i}}^{t_{f}} dt 
V_{e^+ e^-} (\vec R, {\dot{\vec z}}_1, {\dot{\vec z}}_2 ) \,,
\label{eq:2.5}
\end{equation}

\noindent 
where  $\vec R = \vec z_1(t) - \vec z_2 (t)$.  To higher order in the
velocities, $i$ log $W(\Gamma)$ cannot be written in the above form and the
concept of a potential is not defined because of the occurence of radiation.
Eq.(\ref{eq:2.5}) does not include contributions of closed loops of
electron positron pairs to $V_{e^+e^-}$.

The integral (\ref{eq:2.4}) is gaussian and has the value

\begin{equation}
W(\Gamma) = e^{-{ie^{2}\over 2}\int_\Gamma dx^\mu \int_{\Gamma}
dx^{\prime\nu}
{D_{\mu\nu}(x - x')\over \epsilon}},
\label{eq:2.6}
\end{equation}

\noindent 
where $D_{\mu\nu}$ is the free photon propagator. 
Letting $\epsilon = 1$ and expanding $i \log W(\Gamma)$ to second order in the
velocities gives: \cite{darwin}

\begin{equation}
V_{e^+e^-} = - {e^2 \over 4\pi R} + {1\over 2} {e^2 \over 4\pi R} \left[{\dot{\vec z}}_1 
\cdot {\dot{\vec z}}_2 + {({\dot{\vec z}}_1  \cdot \vec R ) ( {\dot{\vec z}}_2 
\cdot \vec R ) \over R^2 }\right] \equiv V_{D} .
\label{eq:2.7}
\end{equation}

\noindent
Furthermore the spin dependent electron positron potential
$V_{e^+e^-}^{\mbox{\scriptsize spin}}$ is determined by the expectation value $\langle \langle
F_{\mu \nu} \rangle \rangle_{Maxwell}$ of the electromagnetic field in the
presence of the external current (\ref{eq:2.1}).

In the dual description first we write the inhomogeneous Maxwell equations in
the form:

\begin{equation}
-\partial^\beta {\epsilon_{\alpha\beta\sigma\lambda} G^{\sigma \lambda}\over
2}
= j_\alpha ,
\label{eq:2.8}
\end{equation}

\noindent
where $G_{\mu\nu}$ is the dual field tensor composed of the electric
displacement vector $\vec D$ and the magnetic field vector $\vec H$:

\begin{equation}
G_{0k} \equiv H_k ,\qquad G_{\ell m} \equiv \epsilon_{\ell mn} D^n.
\label{eq:2.9}
\end{equation}

Next attach a line $L$ of polarization charge between the electron positron
pair.  As the charges move the line $L$ sweeps out a surface  $y^{\alpha}
\left(\sigma, \tau \right)$ bounded by $\Gamma$ (the Dirac
sheet) and generates
the Dirac polarization tensor $G_{\mu \nu}^S \left( x \right)$: \cite{dirac}

\begin{equation}
G_{\mu\nu}^S (x) = - e \epsilon_{\mu\nu \alpha\beta} \int d\sigma \int d \tau
{\partial y^\alpha\over\partial\sigma} {\partial y^\beta\over\partial\tau}
\delta (x - y(\sigma,\tau)).
\label{eq:2.10}
\end{equation}

\noindent 
The current density (\ref{eq:2.1}) can then be written in the form:  \cite{dirac}

\begin{equation}
 - \partial^\beta {\epsilon_{\alpha\beta\sigma\lambda}
G^{S\sigma\lambda}(x)
\over 2}=j_\alpha (x) ,
\label{eq:2.11}
\end{equation}

\noindent 
and the solution of the inhomogeneous Maxwell equations (\ref{eq:2.8}) is

\begin{equation}
G_{\mu\nu} = \partial_\mu C_\nu - \partial_\nu C_\mu + G_{\mu\nu}^S,
\label{eq:2.12}
\end{equation}

\noindent 
which defines the magnetic variables (the dual potentials $C_\mu$).  

The homogeneous Maxwell equations for $\vec E$ and $\vec B$, written in
the form

\begin{equation}
\partial^\alpha (\mu G_{\alpha\beta}) = 0,
\label{eq:2.13}
\end{equation}

\noindent 
where $\mu = {1\over\epsilon}$ is the magnetic susceptibility, become
dynamical equations for the dual potentials, and can be obtained
by varying $C_\mu$ in the Lagrangian

\begin{equation}
{\cal L}_C (G_{\mu\nu}^S) = - {1\over 4} \mu G_{\mu\nu} G^{\mu\nu} \,,
\label{eq:2.14}
\end{equation}

\noindent 
where $G_{\mu\nu}$ is given by (\ref{eq:2.12}).  This Lagrangian
provides the
dual (magnetic) description of the Maxwell theory (\ref{eq:2.2}).
  In the dual description the Wilson loop $W(\Gamma)$
 is  given by

\begin{equation}
W (\Gamma) = {\int {\cal D}C_\mu e^{i\int dx [{\cal L}_C (G_{\mu\nu}^S) +
{\cal L}_{GF}]}\over \int {\cal D} C_\mu e^{i\int dx [{\cal L}_C
(G_{\mu\nu}^S= 0) + {\cal L}_{GF}]}}.
\label{eq:2.15}
\end{equation}

The functional integral (\ref{eq:2.15}) is also Gaussian and has the value (\ref{eq:2.6}) with
$1 \over \epsilon$ replaced by $\mu$.  We then have two equivalent
descriptions at all distances of the electromagnetic interaction of two
charged particles.

Note from (\ref{eq:2.2}) and (\ref{eq:2.14}) that the equations

\begin{equation}
\epsilon= {1 \over g_{el}^2}, \hspace{.25in} 
\mu= {1 \over g_{mag}^2}
\label{eq:2.16}
\end{equation}

\noindent 
define electric and magnetic coupling constants.  If the wave number dependent
dielectric constant $\epsilon \rightarrow 0$ at long distances, then $g_{el}
\rightarrow \infty$ and the Maxwell potentials $A_\mu$ are strongly coupled.
By contrast, $g_{mag} \rightarrow 0$, and the dual potentials are weakly 
coupled at large distances.  

\section{The Heavy Quark Potential in QCD}

The heavy quark potential $V_{q \bar q}$ is determined by the Wilson loop $W
(\Gamma)$ of Yang Mills theory:

\begin{equation}
W (\Gamma) = {\int {\cal D} Ae^{iS_{YM}(A)} tr P\exp (-ie \oint_\Gamma dx^\mu
A_\mu (x))\over \int {\cal D} Ae^{iS_{YM} (A)}}.
\label{eq:3.1}
\end{equation}

\noindent  
(See Fig.1)
   As usual $A_\mu (x) = {1\over 2}
\lambda_a A_\mu^a (x)$, $tr$ means the trace over color indices,  $P$
prescribes the ordering of the color matrices according to the direction fixed
on the loop and $S_{YM}(A)$ is the Yang--Mills action including a gauge fixing
term.  We have denoted the Yang--Mills coupling constant by e, i.e.,

\begin{equation}
\alpha_s = {e^2\over 4\pi}.
\label{eq:3.2}
\end{equation}

The spin independent part $V ( \vec R, {\dot{\vec z}}_1, {\dot{\vec z}}_2)$
of $V_{q \bar q}$ is obtained from (\ref{eq:3.1}) by the QCD analogue of
(\ref{eq:2.5}): 

\begin{equation}
i \log W (\Gamma) = \int_{t_i}^{t_f} dt V (\vec R, {\dot{\vec z}}_1, {\dot{\vec z}}_2 ).
\label{eq:3.3}
\end{equation}

The spin dependent heavy quark potential $V^{\mbox{\scriptsize spin}}$ is a sum of terms
depending upon quark spin matrices, masses, and momenta: \cite{calc}

\begin{equation}
V^{\mbox{\scriptsize spin}}= V_{LS}^{MAG} + V_{Thomas} + V_{Darwin} + V_{SS},
\label{eq:3.4}
\end{equation}

\noindent
where the notation indicates the physical significance of the individual
terms (MAG denotes magnetic).  Each term in (\ref{eq:3.4}) can be obtained from a
corresponding term in $V_{e^+e^-}^{\mbox{\scriptsize spin}}$ by making the replacement

\begin{equation}
\langle \langle F_{\mu \nu} \left(z_1 \right) \rangle \rangle_{Maxwell} \longrightarrow \langle \langle F_{\mu \nu} \left(z_1 \right) \rangle \rangle_{YM} , 
\label{eq:3.5}
\end{equation}

\noindent
where

\begin{equation}
\langle \langle F_{\mu \nu} \left(z_1 \right) \rangle \rangle_{YM} \equiv {\int {\cal D} Ae^{iS_{YM}(A)} tr P \exp [-ie \oint_\Gamma dx^\mu
A_\mu (x)] F_{\mu \nu} \left(z_1 \right)   \over \int {\cal D} Ae^{iS_{YM}(A)} tr P \exp [-ie \oint_
\Gamma dx^{\mu} A_\mu (x)]} ,
\label{eq:3.6}
\end{equation}

\noindent
and

\begin{equation}
F_{\mu\nu} = \partial_\mu A_\nu - \partial_\nu A_\mu - ie [A_\mu, A_\nu],
\label{eq:3.7}
\end{equation}

\noindent
i.e. $\langle\!\langle F_{\mu\nu}(x)\rangle\!\rangle_{YM}$ is the expectation value of
the Yang--Mills field tensor in the presence of a quark and anti--quark moving
along classical trajectories $\vec z_1 (t)$ and $\vec z_2 (t)$ respectively.

The calculation of the heavy quark potential is then reduced to the evaluation
of functional integrals of Yang Mills theory.   Because of the strong coupling
at long distances all field configurations can give important contributions to
(\ref{eq:3.1}) and (\ref{eq:3.6}) for large loops $\Gamma$ and there is no simple description
in terms of Yang Mills potentials.  

\section{The Dual Description of Long Distance Yang-Mills Theory}

The dual theory described here is a concrete realization of the Mandelstam \linebreak 't Hooft \cite{mand} dual
superconductor picture of confinement.  A dual Meissner effect prevents the
electric color flux from spreading out as the distance $R$ between the quark
anti-quark pair increases.  As a result a linear potential develops which
confines the quarks in hadrons.  Such a dual picture is suggested
by the solution of a truncated set of Dyson
equations of Yang Mills theory \cite{nph81}  which gives an effective dielectric constant
$\epsilon(q)\rightarrow q^2/M^2$ as $q^2 \rightarrow 0$ ($M$ is an
undetermined mass scale).  As a consequence $\mu = {1 \over \epsilon}
\rightarrow {M^2 \over q^2}$ as $q^2 \rightarrow 0$ so that the dual gluon
becomes massive as is characteristic of dual superconductivity.  However, such
a truncation cannot be justified in the strongly coupled domain and duality in
Yang Mills theory remains an hypothesis.

On the other hand, there has been a recent revival of interest in
electric-magnetic duality due to the work of Seiberg and Witten \cite{npb94} on
supersymmetric $N=2$ Yang Mills theory and Seiberg \cite{npb95} on $N=1$ supersymmetric
QCD.  The long distance physics of these models, which are asymptotically
free, is described by weakly coupled dual gauge theories.  These examples of
non-Abelian gauge theories for which duality can can be inferred provide new
motivation for the duality hypothesis for Yang Mills theory.

The dual theory is 
described by an effective  Lagrangian density ${\cal L}_{\mbox{\scriptsize eff}}$ in which the fundamental
variables are an octet of dual potentials ${\bf C_{\mu}}$ coupled minimally to
three octets of scalar Higgs fields ${\bf B}_i$ carrying magnetic color
charge \cite{calc,physrev68}.  (The gauge coupling constant of the dual theory $g = {2 \pi \over e}$). The Higgs potential
has a minimum at non-zero values ${\bf B}_{0i}$ which have the color
structure

\begin{equation}
{\bf B}_{01} =  B_0 \lambda_7 ,
\quad {\bf B}_{02} = B_0 (-\lambda_5),\quad {\bf
B}_{03} = B_0 \lambda_2.
\label{eq:4.1}
\end{equation}

\noindent
The three matrices $\lambda_7, - \lambda_5$ and $\lambda_2$ transform as a
$j=1$ irreducible representation of an $SU(2)$ subgroup of $SU(3)$ and as
there is no $SU(3)$ transformation which leaves all three ${\bf B}_{0i}$
invariant the dual $SU(3)$ gauge symmetry is completely broken and the eight
Goldstone bosons become the longitudinal components of the now massive
$\bf{C}_\mu$.

The basic manifestation of the dual superconducting properties of ${\cal L}_{\mbox{\scriptsize eff}}$ is
that it generates classical equations of motion having solutions \cite{physrev90} carrying
a unit of $Z_3$ flux confined in a narrow tube along the $z$ axis
(corresponding to having quark sources at $z = \pm \infty$). (These solutions
are dual to Abrikosov-Nielsen-Olesen magnetic vortex solutions \cite{abrikosov} in a
superconductor).  Before writing ${\cal L}_{\mbox{\scriptsize eff}}$ we briefly describe these classical
solutions. The monopole fields
${\bf B}_i$ have the form : \cite{calc}

\begin{eqnarray}
{\bf B}_1 & = & B_1(x) \lambda_7 + \bar B_1(x)(-\lambda_6)\,, 
\nonumber \\
{\bf B}_2 & = & B_2(x)(-\lambda_5) + \bar B_2(x) \lambda_4 \,,
\label{eq:4.2}
\\
{\bf B}_3 & = & B_3(x)\lambda_2 + \bar B_3(x)(-\lambda_1) \,.
\nonumber
\end{eqnarray}

\noindent
We denote

\begin{equation}
\phi_i(x) = B_i(x) - i \bar B_i(x) \,,
\label{eq:4.3}
\end{equation}

\noindent
and look for solutions where the dual potential is proportional to the
hypercharge matrix $Y = {\lambda_8 \over \sqrt{3}}$,

\begin{equation}
{\bf C}_\mu = C_\mu Y ,
\label{eq:4.4}
\end{equation}

\noindent
and where 
\begin{equation}
\phi_1(x) = \phi_2(x) \equiv \phi(x), \hspace{.25in}
\phi_3(x) = B_3(x).
\label{eq:4.5}
\end{equation}

\noindent
At large distances from the center of the flux tube in cylindrical
coordinates $\rho,\theta,z$ the boundary conditions are:

\begin{equation}
\vec C \rightarrow - {\hat e_\theta \over g \rho},\quad ~\phi \rightarrow
B_0 e^{i\theta},\quad  B_3 \rightarrow B_0, ~\quad \mbox{as} \ ~ \rho \rightarrow \infty .
\label{eq:4.6}
\end{equation}

The non-vanishing of $B_0$ produces a color monopole current confining the
electric color flux.  The line integral of the dual potential around a large
loop surrounding the $z$ axis measures this flux, and the
boundary condition (\ref{eq:4.6}) for $\vec{\bf C}$ gives

\begin{equation}
e^{- ig \oint_{loop} \vec {\bf C} \cdot  d \vec \ell} = e^{2\pi i Y} =
e^{2\pi\left({i\over 3}\right)},
\label{eq:4.7}
\end{equation}

\noindent
which manifests the unit of $Z_3$ flux in the tube.
  The energy per unit
length in this flux tube gives the string
tension $\sigma$: \cite{physrev90}

\begin{equation}
\sigma \sim 24B_0^2  .
\label{eq:4.8}
\end{equation}

The field $\phi (\vec x)$ vanishes at the center of the flux tube.  By contrast
$B_3(\vec{x})$ does not couple to quarks and remains close to its vacuum value
for all $\vec x$.  For simplicity in the rest of this talk we set 
$B_3(x) = B_0$, in which case ${\cal L}_{\mbox{\scriptsize eff}}$ reduces to the Abelian Higgs model.

To couple ${\bf C}_\mu$ to a $q \bar q$ pair separated by a finite distance we
represent quark sources by a Dirac string tensor ${\bf G}_{\mu \nu}^S$.  We
choose the dual potential to have the same color structure (\ref{eq:4.4}) as the
flux tube solution.  Then ${\bf G}_{\mu \nu}^S$ must
also be proportional to the hypercharge matrix,

\begin{equation}
{\bf G}_{\mu \nu}^S = YG_{\mu \nu}^S ,
\label{eq:4.9}
\end{equation}

\noindent
where $G_{\mu \nu}^S$ is given by (\ref{eq:2.10}), so that one unit of $Z_3$ flux
flows along the Dirac string connecting the quark and anti-quark. With the
ansaetze (\ref{eq:4.9}) and (\ref{eq:4.2})- (\ref{eq:4.5}) along with the simplification 
$B_3(x)=B_0$ , the Lagrangian density ${\cal L}_{\mbox{\scriptsize eff}}\left( G_{\mu \nu}^S \right)$ coupling
dual potentials to classical quark sources moving along trajectories $\vec
z_1(t)$ and $\vec z_2(t)$ assumes the form:

\begin{equation}
{\cal L}_{\mbox{\scriptsize eff}} ( G_{\mu \nu}^S ) = - {4 \over 3} {\left( G_{\mu \nu}G^{\mu \nu} \right) \over 4} + { 8 | (\partial_\mu - igC_\mu) \phi |^2 \over 2} - {100 \over 3} \lambda \left( |\phi |^2 - B_0^2 \right)^2 ,
 \label{eq:4.10}
\end{equation}

\noindent
where
\begin{equation}
G_{\mu \nu} = \partial_\mu C_\nu - \partial_\nu C_\mu + G_{\mu \nu}^S \,,
\label{eq:4.11}
\end{equation}

\noindent
and 

\begin{equation}
g= {2 \pi \over e}.
\label{eq:4.12}
\end{equation}

The first term in ${\cal L}_{\mbox{\scriptsize eff}}$ is the coupling of dual potentials to
quarks, the second is the coupling of the dual potentials to monopole fields
$\phi$, while the third term is the quartic self coupling of the monopole
fields.  The numerical factors in (\ref{eq:4.10}) arise from inserting the color
structures (\ref{eq:4.2})- (\ref{eq:4.5}) in the original non-Abelian form of ${\cal L}_{\mbox{\scriptsize eff}}$.
By a suitable redefinition of $\phi$ and $\lambda$ the last two terms can be
written in the standard form of the Abelian Higgs model, while the color
factor $4 \over 3$ in the first term is a consequence of (\ref{eq:4.4}) and (\ref{eq:4.9}),
which combined with the boundary condition (\ref{eq:4.6}) provides the unit of $Z_3$
flux.

We find from (\ref{eq:4.10}) the following values of the dual gluon mass $M$ and the
monopole mass $M_\phi$ :

\begin{equation}
M^2 = 6g^2B_0^2 \hspace{.15in}, \hspace{.25in}  M_\phi^2 = {100 \lambda \over 3} B_0^2 .
\label{eq:4.13}
\end{equation}

\noindent
The quantity $g^2/\lambda$ plays the role of a Landau-Ginzburg parameter.  Its
value can be estimated by relating the difference between the energy density
at a large distance from the flux tube and the energy density at its center to
the gluon condensate.\cite{physrev90} This procedure gives $g^2/\lambda \simeq  5$. There  
remain two free parameters in ${\cal L}_{\mbox{\scriptsize eff}}$, which we take to be
$\alpha_s = { e^2 \over 4 \pi} = {\pi \over g^2 }$ and the string tension
$\sigma$. 

We denote by $W_{\mbox{\scriptsize eff}}(\Gamma)$ the Wilson loop of the dual theory,
i.e.,

\begin{equation}
  W_{\mbox{\scriptsize eff}} (\Gamma) =
   {
   \int {\cal D} C_\mu {\cal D} \phi 
    e ^ {i \int dx [ {\cal L}_{\mbox{\scriptsize eff}} (G_{\mu\nu}^S) + {\cal L}_{GF} ] }
  \over
  \int {\cal D} C_\mu {\cal D} \phi 
    e ^ {i \int dx [ {\cal L}_{\mbox{\scriptsize eff}} (G_{\mu\nu}^S=0) + {\cal L}_{GF} ] }
   }.
  \label{eq:4.14}
\end{equation}

\noindent
\noindent
The functional integral $W_{\mbox{\scriptsize eff}}(\Gamma)$ determines in the effective dual
theory the same physical quantity as $W(\Gamma)$ in Yang-Mills theory, namely
the action for a quark anti-quark pair moving along classical trajectories.
The coupling of dual potentials to Dirac strings in ${\cal L}_{\mbox{\scriptsize eff}} \left(G_{\mu \nu}^S \right)$ 
  plays the role in eq.(\ref{eq:4.14}) for
$W_{\mbox{\scriptsize eff}}(\Gamma)$ of the Wilson loop $Pe^{-ie \oint_\Gamma dx^\mu
A_\mu(x)}$ in eq.(\ref{eq:3.1}) for $W(\Gamma)$.

The assumption that the dual theory describes the long distance $q \bar q$
interaction in Yang-Mills theory then takes the form:

\begin{equation}
W(\Gamma) = W_{\mbox{\scriptsize eff}}(\Gamma), ~{\rm for ~large ~loops ~\Gamma}.
\label{eq:4.15}
\end{equation}

\noindent
Large loops mean that the size $R$ of the loop is large compared to the
inverse of $M$ and $M_\phi$.  Since the dual theory is weakly
coupled at large distances we can evaluate $W_{\mbox{\scriptsize eff}}(\Gamma)$ via a
semi-classical expansion to which the classical configuration of dual
potentials and monopoles gives the leading contribution.  Furthermore using (\ref{eq:4.15}), we can relate the
expectation value (\ref{eq:3.6}) of the Yang Mills Field tensor at the position of a quark to the corresponding
expectation value of the dual field tensor in the effective theory: \cite{calc}

\begin{equation}
\langle \langle F_{\mu\nu}(z_1)\rangle\rangle_{YM} ={4 \over 3} \langle
\langle \hat G_{\mu \nu}(z_1) \rangle\rangle_{\mbox{\scriptsize eff}}  ,
\label{eq:4.16}
\end{equation}

\noindent
where
\begin{equation}
\hat G_{\mu\nu}(x) \equiv {1 \over 2} \epsilon_{\mu \nu \lambda \sigma}
G^{\lambda \sigma}(x),
\label{eq:4.17}
\end{equation}

\noindent
and
\begin{equation}
\langle\!\langle G^{\mu \nu}(z_1)\rangle\!\rangle_{\mbox{\scriptsize eff}} \equiv {\int {\cal
D}
C_\mu {\cal D} \phi e^{i\int dx
({\cal L}_{\mbox{\scriptsize eff}} (G_{\mu\nu}^S) + {\cal L}_{GF})} G^{\mu \nu} (z_1) \over 
\int
{\cal D} C_\mu {\cal D} \phi  e^{i\int dx
 ({\cal L}_{\mbox{\scriptsize eff}}
(G_{\mu\nu}^S) + {\cal L}_{GF})}}.
\label{eq:4.18}
\end{equation}

To obtain the spin independent heavy quark potential $V({\vec R},{\dot{\vec z}}_1, {\dot{\vec z}}_2
)$ in the dual theory we replace $W(\Gamma)$ by $W_{\mbox{\scriptsize eff}}(\Gamma)$ in eq.(\ref{eq:3.3}).  This
expresses the spin independent heavy quark potential in terms of the zero order and quadratic
terms in the expansion of $i \log W_{\mbox{\scriptsize eff}}(\Gamma)$ for small velocities ${\dot{\vec z}}_1$ and ${\dot{\vec z}}_2$.
The corresponding spin dependent potential in the dual theory is obtained by
making the replacement

\begin{equation}
\langle \langle F_{\mu \nu} (z_1) \rangle \rangle_{YM} \longrightarrow \hspace{.15in}
{4 \over 3} \langle \langle \hat
G_{\mu \nu} (z_1) \rangle \rangle_{\mbox{\scriptsize eff}} ,
\label{eq:4.19}
\end{equation}

\noindent
in the expressions in eq.(\ref{eq:3.4}) for $V^{\mbox{\scriptsize spin}}$.

\section{The Classical Approximation to the Dual Theory}
In the classical approximation all quantities are replaced by their classical
values

\begin{equation}
\langle\langle G_{\mu \nu}(x) \rangle\rangle_{\mbox{\scriptsize eff}} = G_{\mu \nu} (x),\hspace{.25in} i \log W_{\mbox{\scriptsize eff}} = - \int dx {\cal L}_{\mbox{\scriptsize eff}}(G_{\mu \nu}^S ),
\label{eq:5.1}
\end{equation}

\noindent
where $G_{\mu \nu}$ and ${\cal L}_{\mbox{\scriptsize eff}}\left(G_{\mu \nu}^S \right)$  are
evaluated at the solution of the classical equations of motion:

\begin{equation}
\partial^\alpha \left(\partial_\alpha C_\beta - \partial_\beta C_\alpha
\right) = - \partial^\alpha G_{\alpha \beta}^S + j_\beta ^{MON},
\label{eq:5.2}
\end{equation}

\begin{equation}
\left( \partial_\mu - igC_\mu \right)^2 \phi = - {200 \lambda \over 3} \phi
\left(|\phi |^2 - B_0^2 \right) ,
\label{eq:5.3}
\end{equation}

\noindent
where the monopole current $j_{\mu}^{MON}$ is

\begin{equation}
j_{\mu}^{MON} = - 3ig[\phi^* \left(\partial_\mu -
igC_\mu \right) \phi - \phi \left(\partial_\mu + igC_\mu \right) \phi^*].
\label{eq:5.4}
\end{equation}

\noindent
The boundary conditions on $\phi$ are:

\begin{equation}
\phi(x) \rightarrow 0, \quad \mbox{as} \  x \rightarrow y\left(\sigma , \tau \right); \qquad
\phi(x) \rightarrow B_0 \,, \quad \mbox{as} \  x \rightarrow \infty.
\label{eq:5.5}
\end{equation}

\noindent
The vanishing of $\phi(x)$  on the Dirac sheet $y^{\mu}(\sigma,\tau)$
produces a flux tube with energy concentrated in the
neighborhood of the string connecting the quark anti-quark pair. Using the minimum energy solution corresponding to a straight
line string , we evaluate $i
\log W_{\mbox{\scriptsize eff}}$ to second order in the velocities ${\dot{\vec z}}_1,$ and
${\dot{\vec z}}_2$ and obtain
 the spin independent heavy quark
potential. 

 At large separations $V (\vec R, {\dot{\vec z}}_1, {\dot{\vec z}}_2 )$ is linear
in $R$ since the monopole current screens the color field of the quarks
so that a color electric Abrikosov-Nielsen-Olesen vortex forms between
the moving $q \bar q$ pair.  For the case of circular motion,
$({\dot{\vec z}}_i \cdot \vec R = 0$, 
$ {\dot{\vec z}}_2 = - {\dot{\vec z}}_1 )$, we find:

\begin{equation}
V \rightarrow \sigma R 
\left[ 1 - A { (\dot{\vec z_1} \times \vec R)^2 \over R^2} \right] \,, 
\quad \mbox{as} \ R \rightarrow \infty \,,
\label{eq:5.6}
\end{equation}

\noindent
where

\begin{equation}
A \simeq .21 \sigma .
\label{eq:5.7}
\end{equation}

\noindent
The constant $A$ determines the long distance moment of inertia $I(R)$ of the
rotating flux tube:

\begin{equation}
\lim_{R\rightarrow\infty} I(R) = {1\over 2} (AR)R^2.
\label{eq:5.8}
\end{equation}

At small separations the color field generated by the quarks expels the
monopole condensate from the region between them and as
$R \rightarrow 0$, $V$ approaches the one gluon
exchange result, ${4 \over 3} V_D$. See eq.(\ref{eq:2.7}).

As the simplest application of this potential, we add relativistic kinetic
energy terms to obtain a classical Lagrangian, and calculate classically the energy and angular momentum of $q \bar q$ circular orbits, which are those which have the largest angular momentum $J$ for a given energy.  We find \cite{95proc} a Regge trajectory $J$ as a function of $E^2$ which for large $E^2$ becomes linear with slope $\alpha^\prime = J/E^2 = 1/8\sigma \left(1-A/\sigma \right)$.  Then (\ref{eq:5.7}) gives $\alpha^\prime \approx 1/6.3 \sigma$, which is close to the string model relation $\alpha^\prime = {1 \over 2 \pi \sigma}$.  This comparison shows how at the classical level a string model emerges when the velocity dependence of the $q \bar q$ potential is included.

  To calculate the spin dependent heavy quark potential we use (\ref{eq:4.19}) and (\ref{eq:5.1}) to evaluate $V^{\mbox{\scriptsize spin}}$ 
(\ref{eq:3.4}) in the classical approximation to the dual theory.
The resulting expressions are given in reference 1. Here we discuss
 only the result for the spin-spin interaction $V_{SS}^{\mbox{\scriptsize spin}}$ between the color magnetic moments of the quark anti-quark pair. 
This magnetic dipole interaction is determined by the gradient of the Greens function $G \left( \vec x, \vec x^\prime \right)$ describing the interaction 
of monopoles: 
\begin{equation}
V_{SS}^{\mbox{\scriptsize spin}} =  {4\over 3} {e^2\over m_1 m_2}
\Bigg\{ (\vec S_1 \cdot \vec S_2)
\delta (\vec z_1 - \vec z_2) - {(\vec S_1 \cdot \vec\nabla)
(\vec S_2 \cdot \vec \nabla^\prime)}  G (\vec x, \vec x^\prime)
\Bigg|_{\vec x= \vec z_1, \vec x^\prime= \vec z_2 }\Bigg\}.
\label{eq:5.9}
\end{equation}
$G$ satisfies the following equation obtained from eq.(\ref{eq:5.2}) for $C^0$ :

\begin{equation}
[-\bigtriangledown^2 + 6g^2 \phi^2(\vec x)]G \left(\vec x, \vec x^\prime \right) = \delta \left( \vec{x} - \vec{x}^\prime \right)\,,
\label{eq:5.10}
\end{equation}

\noindent
where $\phi (\vec x)$ is the static monopole field. ($\phi(x)$ is real so that the monopole charge density $j^0(x) = 6g^2
\phi^2(x)C^0$.)  Since $\phi (\vec x)$ approaches its vacuum value $B_0$ as $\vec x
\rightarrow \infty$, $G$ vanishes exponentially at large distances:

\begin{equation}
G(\vec x, \vec x')
\mathrel {\mathop {\longrightarrow}_{\vec x \to \infty}}
{e^{-M |\vec x - \vec x'|}\over 4\pi |\vec x - \vec x'|}\,.
\label{eq:5.11}
\end{equation}

\noindent
See eq.(\ref{eq:4.13}).

The dual Higgs mechanism then produces the long distance
Yukawa potential (\ref{eq:5.11})  between
monopoles along with the linear potential (\ref{eq:5.6}) between quarks.  The resulting
suppression of the color magnetic interaction between quarks is an unambiguous
prediction of electric-magnetic duality. 

\section{Fluctuations of the Flux Tube and Effective String Theory}

To evaluate the contributions to $W_{\mbox{\scriptsize eff}}$ arising
from fluctuations of the shape and length of the flux tube we must
integrate over field configurations generated by all strings connecting
the $q \bar q$ pair.  This amounts to doing a functional integral over
all polarization tensors $G_{\mu \nu}^S(x)$.  Similar integrals have
recently been carried out by Akhmedov et al. \cite{string} in the case
$\lambda \rightarrow \infty$.  By changing from field variables to
string variables ,the functional integral over $G_{\mu \nu}^S(x)$ is
replaced by a functional integral over corresponding world sheets
$y^\mu \left(\sigma,\tau \right)$, multipled by an appropriate Jacobian
and there results \cite{string}  an effective string theory free from
the conformal anomaly \cite{plb81} Such techniques if extended to
finite $\lambda $ could be applied to $W_{\mbox{\scriptsize eff}}$ to
obtain a corresponding effective string theory . The leading long
distance contribution to the static potential due to fluctuations of
the string which is independent of the details of the string theory
would then have the universal value $ - {\pi \over 12R} $.
\cite{npb81}

\section{Conclusion} 

We have obtained an expression for the heavy quark potential $V_{q\bar
q}$ in terms of an effective Wilson loop $W_{\mbox{\scriptsize eff}}
(\Gamma)$ determined by the dynamics of a dual theory which is weakly
coupled at long distances.  The classical approximation gives the
leading long distance contribution to $W_{\mbox{\scriptsize
eff}}(\Gamma)$ and yields a velocity dependent spin dependent heavy
quark potential which for large $R$ becomes linear in $R$ and which for
small $R$ approaches lowest order perturbative QCD. The dual theory
cannot describe QCD at shorter distances, where radiative corrections
giving rise to asymptotic freedom become important. At such distances
the dual potentials are strongly coupled and the dual description is no
longer appropriate.  

As a final remark we note that the dual theory is an $SU(3)$ gauge
theory, like the original Yang-Mills gauge theory.  However, the
coupling to quarks selected out only Abelian configurations of the dual
potential.  Therefore, our results for the $q\bar q$ interaction do not
depend upon the details of the dual gauge group and should be regarded
more as consequences of the general dual superconductor picture rather
than of our particular realization of it.

\section{Acknowledgments}

I would like to thank N. Brambilla and G. M. Prosperi for the opportunity to
 attend this conference  and for their important contributions	
to the work presented here.

\end{document}